\documentclass[conference]{IEEEtran}

\usepackage{amsmath,amssymb,amsfonts}
\usepackage{algorithmic}
\usepackage{graphicx}
\usepackage{textcomp}
\usepackage{xcolor}

\begin{document}

\title{Multiscale Latent Diffusion Model for Enhanced Feature Extraction from Medical Images}

\author{\IEEEauthorblockN{1\textsuperscript{st} Rabeya Tus Sadia}
\IEEEauthorblockA{\textit{dept. of Computer Science} \\
\textit{University of Kentucky}\\
Kentucky, USA \\
rabeya.sadia@uky.edu}
\and
\IEEEauthorblockN{2\textsuperscript{nd} Jie Zhang}
\IEEEauthorblockA{\textit{Department of Radiology} \\
\textit{University of Kentucky}\\
Kentucky, USA \\
jnzh222@uky.edu}
\and
\IEEEauthorblockN{3\textsuperscript{rd} Jin Chen}
\IEEEauthorblockA{\textit{Department of Medicine-Nephrology} \\
\textit{University of Alabama at Birmingham}\\
Birmingham, USA \\
jinchen@uab.edu}
}

\maketitle

\begin{abstract}

Various imaging modalities are used in patient diagnosis, each offering unique advantages and valuable insights into anatomy and pathology. Computed Tomography (CT) is crucial in diagnostics, providing high-resolution images for precise internal organ visualization. CT's ability to detect subtle tissue variations is vital for diagnosing diseases like lung cancer, enabling early detection and accurate tumor assessment. However, variations in CT scanner models and acquisition protocols introduce significant variability in the extracted radiomic features, even when imaging the same patient. This variability poses considerable challenges for downstream research and clinical analysis, which depend on consistent and reliable feature extraction. Current methods for medical image feature extraction, often based on supervised learning approaches, including GAN-based models, face limitations in generalizing across different imaging environments. In response to these challenges, we propose LTDiff++, a multiscale latent diffusion model designed to enhance feature extraction in medical imaging. The model addresses variability by standardizing non-uniform distributions in the latent space, improving feature consistency. LTDiff++ utilizes a UNet++ encoder-decoder architecture coupled with a conditional Denoising Diffusion Probabilistic Model (DDPM) at the latent bottleneck to achieve robust feature extraction and standardization. Extensive empirical evaluations on both patient and phantom CT datasets demonstrate significant improvements in image standardization, with higher Concordance Correlation Coefficients (CCC) across multiple radiomic feature categories. Through these advancements, LTDiff++ represents a promising solution for overcoming the inherent variability in medical imaging data, offering improved reliability and accuracy in feature extraction processes.
\end{abstract}
\section{Introduction}
One of the most prevalent malignancies affecting both men and women in the US, lung cancer continues to be the leading cause of cancer-related mortality \cite{collins2016letter}. Non-small cell lung cancer (NSCLC) has a five-year survival rate of about 19\%. In order to improve treatment outcomes, computed tomography (CT) imaging is crucial for the early diagnosis of lung cancer and for defining the features of the tumor \cite{de2014benefits, ravanelli2013texture}. Additionally, texture analysis of CT images makes it easier to quantify temporal and geographical variations in the functionality and structure of tumors, which allows for the evaluation of intra-tumor evolution \cite{ardila2019end},\cite{song2017using}. Image analysis technology is utilized to extract clinically relevant information from medical images, while image processing techniques enable the efficient and accurate segmentation of key features. Together, these technologies form an essential foundation for physicians in developing consultation strategies, surgical interventions, and diagnostic plans \cite{yan2024survival}. Moreover, due to the complexity, variability, and individual differences in organ structures within the human body, effectively extracting relevant features from medical images is critical for enhancing the accuracy of disease diagnosis and treatment planning. Currently, the most widely utilized techniques include the Gaussian curve method, direction template method, matrix method, and threshold method. Some researchers have introduced a novel approach for subpixel boundary extraction using Gaussian fitting. This approach involves selecting a set of points along the boundary of the original image, applying grayscale processing to those points, and then performing high-precision fitting for accurate boundary extraction \cite{hu2024exploration}, \cite{sadia2024ct}.\\
Feature extraction from CT images plays a pivotal role in medical imaging tasks, including segmentation, classification, and disease diagnosis. Traditional methods for feature extraction often relied on handcrafted techniques, but deep learning algorithms have revolutionized this field by automatically learning and extracting hierarchical features from raw images \cite{lecun2015deep}. Convolutional Neural Networks (CNNs) have become the backbone of many medical imaging applications due to their ability to capture spatial hierarchies in pixel data. In particular, CNNs can be fine-tuned to recognize specific patterns, such as lesions or tumors, which may be critical for early diagnosis and treatment planning \cite{litjens2017survey}. These models excel at capturing local features, such as edges and textures, while progressively learning more abstract representations in deeper layers.

Recently, diffusion models have emerged as a powerful generative approach for enhancing image quality and extracting subtle features that might be overlooked by traditional CNNs. Diffusion models are based on iterative denoising processes, where a noisy image is progressively refined to recover detailed structures \cite{ho2020denoising}. This technique is particularly useful in medical imaging, where noise and low contrast often obscure important features in CT scans. For example, Denoising Diffusion Probabilistic Models (DDPMs) have been shown to effectively reconstruct high-quality images from noisy inputs, preserving fine details such as the boundaries of tumors or lesions \cite{nichol2021improved}. By enhancing the visibility of these critical regions, diffusion models provide more informative features for downstream tasks, such as classification and segmentation.

Combining deep learning approaches with diffusion models offers a promising direction for medical imaging. CNNs provide a solid framework for feature extraction, while diffusion models augment the process by refining and enhancing these features, especially in challenging cases where noise or artifacts may degrade image quality. In particular, diffusion models can improve the clarity of regions of interest (ROI), such as tumors, which are often difficult to delineate in raw CT images. This fusion of techniques enables a more comprehensive analysis of the medical data, improving both the accuracy of diagnostic models and the interpretability of results. By leveraging the strengths of both deep learning and diffusion models, researchers are advancing the state-of-the-art in medical image analysis, providing more robust tools for clinical applications \cite{shen2023improving}.
In comparison to established generative models such as Generative Adversarial Networks (GANs) and Variational Autoencoders (VAEs), Denoising Diffusion Probabilistic Models (DDPMs) have demonstrated superior performance in image standardization tasks. DDPMs use a Markov chain process that progressively transforms a simple starting distribution, like isotropic Gaussian noise, into the target data distribution. This approach involves two key stages: first, a forward diffusion process gradually adds noise to the image by sequentially sampling latent variables, and second, a reverse process uses a neural network, often a U-Net, to progressively remove the noise, recovering a clear image from the noisy input. Since gaining attention in 2020, DDPMs and their variants have driven major advancements in data modeling, with notable successes in areas like image generation, super-resolution, and image-to-image translation. More recently, conditional DDPMs have shown exceptional results in generating images based on specific conditions, while latent DDPMs have enabled efficient image generation within a low-dimensional latent space.

The training process for our proposed model, LTDiff++ involves three main phases. First, the multiscale UNet++ encoder-decoder network is trained using all available CT images, whether standardized or not, to learn how to encode each image into a latent vector that can be used to reconstruct the original image with minimal information loss. Next, a latent conditional DDPM is trained on pairs of non-standard and standardized images, allowing the model to learn the conditional probability distribution required to synthesize standardized images. Finally, the trained networks are combined to get the enhanced features from new CT images, enabling the robust harmonization of image data while maintaining crucial structural features.
\section{Background}
\subsection{ CT Image Acquisition and Reconstruction Parameters}
CT images are obtained by precisely adjusting several acquisition parameters, such as kilovoltage peak (kVp), pitch, milliampere-seconds (mAs), reconstruction field of view (FOV), slice thickness, and reconstruction kernels. Variations in CT image capture and reconstruction parameters, together with the use of different CT scanners, can drastically alter the radiomic features extracted from the pictures. For example, as seen in Figure 1, the Br40 kernel helps to provide a more homogeneous image texture, but the Bl64 kernel improves image sharpness. These textural variations are important because they result in the extraction of diverse radiomic features, which complicates subsequent clinical evaluations and decision-making processes. 

\begin{figure}[htbp]
  \centering
   \includegraphics[angle=180, scale=-1, width=0.5\textwidth]{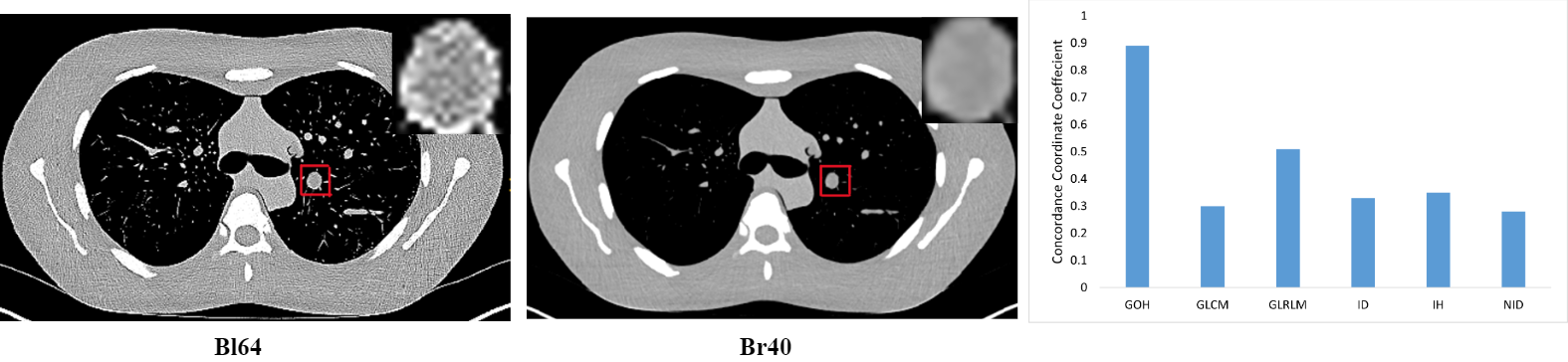}
  \caption{Shows how disparities in imaging procedures might result in differences in tumor picture characteristics. The same scanner was used to scan the identical lungman chest phantom. Two distinct image reconstruction techniques were utilized to get CT images kernels appropriately, as the text at the bottom of the pictures shows. Red squares are used to indicate tumors in the images on the left(tumor zoomed in the top left corner). The feature variance in terms of CCC between these two tumors was displayed by the histogram on the right. The potential of extensive radiomic features may be significantly impacted by the detected variations in the tumor pictures.}
\end{figure}
\subsection{Radiomic Features}
Radiology uses cutting-edge non-invasive imaging techniques to identify and treat a variety of illnesses. Accurate tumor characterization depends on the image attributes that are derived from radiological images using statistical and mathematical models \cite{yip2016applications}. Within the field of medical imaging analysis, radiomic features provide deep insights into the genetic and cellular aspects of phenotypic patterns that are unseen to the human eye \cite{yip2016applications, basu2011evolving, yang2011quantifying}. The following six categories can be used to systematically classify these features in order to aid in a deeper understanding of underlying biological processes: Neighbor Intensity Difference (NID), Gradient Oriented Histogram (GOH), Gray Level Co-occurrence Matrix (GLCM), Gray Level Run Length Matrix (GLRLM), Intensity Direct (ID), and Intensity Histogram (IH). In clinical diagnostic and research situations, these classifications improve the interpretability and application of radiomic data.
\subsection{Existing Methods}
CT image standardization can be broadly divided into two categories, each based on data availability and customized to meet particular needs. The first method requires the existence of paired image data and is called intra-scanner image standardization. By comparing and correcting images taken with varied settings, this method is crucial for standardizing images within the same scanner \cite{selim2020stan}.\\
Each image pair in this method consists of two images from the same scan that have been processed using different reconstruction kernels; the target image utilizes a standard kernel (e.g., Siemens Bl64) and the source image is created using a non-standard kernel (e.g., Siemens Br40). This paired image data is used by a machine learning model to train itself to convert source images into matching target images. Instead of depending on paired data, cross-scanner standardization models are used in the second type of CT image standardization technique. Images are not required to correlate in this paradigm directly; instead, they are acquired independently using different protocols and stored in various locations.
\section{Method} 
Figure 2 illustrates the architecture of the multiscale latent Diffusion model, which integrates with primary elements: a multiscale residual-based image embedding mechanism and a conditional Deep Diffusion Probabilistic Model (DDPM) operating within the latent space under deep supervision. The image embedding mechanism utilizes a residual-based encoder-decoder network to convert input Computed Tomography (CT) images into a compact, low-dimensional latent representation. Following this, the conditional DDPM is employed to model the conditional probability distribution of the latent representations, enabling the generation of standardized images. A significant characteristic of the multiscale model lies in its utilization of a latent diffusion network underpinned by deep supervision. Within the UNet architecture, the encoder and decoder sub-networks are interconnected via a sequence of intricately nested, dense skip pathways. These reconfigured skip pathways are strategically designed to diminish the semantic discrepancy between the feature maps of the sub-networks, thereby simplifying the learning process. 
\begin{figure*}[htbp]
  \centering
   \includegraphics[angle=180, scale=-1, width=0.8\textwidth]{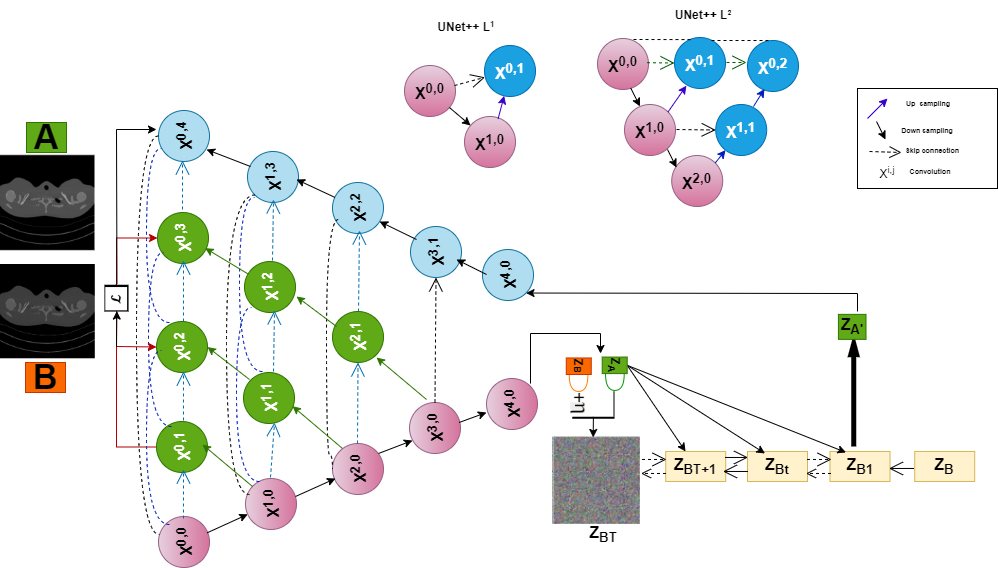}
  \caption{ \textbf{Overview of LTDiff++ architechture. }
     \(A\) and \(Z_A\): non-standard image and its latent vector.
    \(B\) and \(Z_B\): standard image and its latent vector.
    \(A'\): standardized image of \(A\) that falls in the \(B\) domain. \(\eta\): Gaussian noise.
    \noindent Given an image pair \((A, B)\) where \(A\) and \(B\) are non-standard and the corresponding standard images, the model aims to synthesize a new image \(A'\) in the domain \(B\). The representation learning component leverages a modified UNet++ encoder-decoder structure. This framework is pivotal for learning encoded latent representations of CT images. Concurrently, the target-specific latent-space mapping component is purpose-built for standard image synthesis. It integrates a DDPM model for effective latent space mapping. Here, \(Z_A\) represents the latent vector of the non-standard image \(A\), \(Z_B\) is the latent vector of the standard image \(B\), \(Z_{A'}\) is the standardized latent vector of image \(A\), and \(\eta\) denotes Gaussian noise.
 }
  \label{}
\end{figure*}
The training process for the model unfolds in three distinct phases. Initially, the multiscale encoder-decoder network, which enhances learning efficiency by facilitating gradient flow, undergoes training with a comprehensive set of CT images from the training dataset, without discriminating based on their standard or non-standard status, or their originating imaging technology, be it GE or Siemens. The objective of this phase is to efficiently translate images into one-dimensional latent vectors, alleviating the vanishing gradient issue, and improving model performance by achieving a reconstruction of the original images with minimal loss of information. In the subsequent phase, the latent conditional DDPM is trained using pairs of images, each pair comprising a non-standard image and its standard counterpart. The multiscale training with deep supervision phase empowers the DDPM to accurately model the conditional probability distribution of the latent representations, thereby facilitating the generation of standardized and enhanced featured images. The final phase involves the integration of all the trained neural networks, optimizing them to standardize new images effectively.
\subsection{Multiscale Encoder Decoder Training}

\subsubsection{UNet++}
We employed UNet++ \cite{zhou2018unet++}, a deeply supervised encoder-decoder architecture where the encoder and decoder sub-networks are linked via nested, dense skip pathways. These redesigned skip pathways are intended to minimize the semantic gap between the feature maps of the encoder and decoder sub-networks. UNet++ is constructed with an encoder and decoder that are interconnected through a series of nested, dense convolutional blocks. The primary objective of UNet++ is to close the semantic gap between the encoder and decoder feature maps before their fusion. For instance, the semantic disparity between \(X_{0,0}\) and \(X_{1,3}\) is addressed using a dense convolution block comprising three convolution layers.

In Fig. 2, black represents the original U-Net, while green and blue illustrate dense convolution blocks along the skip pathways, and red highlights deep supervision. The colors red, green, and blue distinctly mark UNet++ from the traditional U-Net. Furthermore, UNet++ can be pruned at inference time if it has been trained under deep supervision.

\subsubsection{Re-designed skip pathways}
In the UNet++ architecture, the skip pathways have been redesigned to enhance the connectivity between the encoder and decoder sub-networks by employing nested, dense convolutional blocks. This redesign aims to bridge the semantic gap between the feature maps of the encoder and decoder, facilitating smoother feature integration prior to fusion. For instance, the skip pathway that connects \(X_{0,0}\) and \(X_{1,3}\) includes a dense convolution block consisting of three convolution layers. Each layer in this block fuses the output from the previous convolution layer with the up-sampled output from a lower block, progressively enriching the feature maps to match the semantic level of the decoder's feature maps. Mathematically, the feature map \(X_{i,j}\) in a nested skip pathway is calculated as follows:

\[
X_{i,j} = 
\begin{cases} 
H(X_{i-1,j}) & \text{if } j = 0 \\
H\left(\left[ X_{i,j-1}, \{ U(X_{i+1,j-1}) \}_{k=0}^{j-1} \right]\right) & \text{if } j > 0
\end{cases}
\]

where \(H\) denotes a convolution followed by activation, \(U\) denotes up-sampling, and the brackets denote the concatenation of feature maps.

\subsubsection{Deep supervision}
Deep supervision in UNet++ is implemented by attaching auxiliary segmentation outputs to each of the decoder layers. These outputs are directly connected to the loss function, enabling early and accurate gradient propagation and helping the model to learn detailed segmentation at multiple semantic levels. The loss function combining these multiple outputs is a weighted sum of the losses computed at each level, potentially using a combination of binary cross-entropy and Dice loss:

\[
L = \sum_{j} w_j \cdot \left(-\frac{1}{N} \sum_{b=1}^{N} \left[ \frac{1}{2} Y_b \log(\hat{Y}_{b,j}) + \frac{2 Y_b \hat{Y}_{b,j}}{Y_b + \hat{Y}_{b,j}} \right] \right)
\]

where \(Y_b\) is the ground truth for the \(b^{th}\) image, \(\hat{Y}_{b,j}\) is the predicted output from the \(j^{th}\) deep supervision layer, \(w_j\) is the weight for the \(j^{th}\) layer's loss contribution, and \(N\) is the number of training samples.

\subsection{Conditional Latent Diffusion Model for Image Standardization}

\subsection*{1. Feature Extraction and Encoding}

Both the non-standard image \(A\) and the standard image \(B\) are encoded into their respective latent representations \(Z_A\) and \(Z_B\). This encoding is performed using a neural network based on the UNet++ architecture, which acts as the encoder \(E\):

\[
Z_A = E(A), \quad Z_B = E(B)
\]

where \(E(\cdot)\) denotes the encoding function that maps an image to its latent representation.

\subsection*{2. Adding Gaussian Noise}

After obtaining the latent representation \(Z_A\) of the non-standard image, Gaussian noise \(\eta\) is added to this representation to initiate the diffusion process:

\[
\tilde{Z}_A = Z_A + \eta, \quad \eta \sim \mathcal{N}(0, \sigma^2 I)
\]

The addition of noise helps simulate the forward diffusion process, where data gradually transitions from a meaningful state to a pure noise state. The noise level \(\sigma^2\) controls the intensity of the noise added.

\subsection*{3. Conditional Diffusion Process}

The core of this methodology is the reverse diffusion process, which is conditioned on the latent space representation \(Z_B\) of the standard image. The purpose is to reverse the noise addition by gradually denoising \(\tilde{Z}_A\) back to a state that is not just a denoised version of \(Z_A\) but also aligned with \(Z_B\):

\[
p_\theta(Z_{A,t-1} | Z_{A,t}, Z_B) = \mathcal{N}(Z_{A,t-1}; \mu_\theta(Z_{A,t}, Z_B, t), \sigma_t^2 I)
\]

In this equation:
\begin{itemize}
    \item \(Z_{A,t}\) denotes the latent state at diffusion time step \(t\).
    \item \(\mu_\theta\) is the parameterized mean function which is learned during training and is conditioned on both \(Z_{A,t}\) and \(Z_B\).
    \item \(\sigma_t^2\) represents the variance at each step, which can either be learned or preset as part of the model's hyperparameters.
\end{itemize}

The reverse process involves iteratively calculating \(Z_{A,t-1}\) from \(Z_{A,t}\) using the above Gaussian distribution, starting from \(Z_{A,T}\) (which is \(\tilde{Z}_A\)) and moving backwards to \(Z_{A,0}\).

\subsection*{4. Decoding to Image Space}

Once the reverse diffusion process is completed, the output \(Z_{A,0}\) is expected to be the denoised and standardized latent representation. This is then decoded back into the image space using a decoder \(D\), which is typically the inverse function of the encoder \(E\):

\[
A' = D(Z_{A,0})
\]

\subsection{Model Training Procedure}

The multuscale conditional Diffusion model is trained in a multi-phase approach that ensures each component is optimized to contribute effectively to the final standardization task. The training phases are as follows:

\begin{enumerate}
    \item \textbf{Training the Encoder-Decoder Network:} Initially, the encoder-decoder framework is trained without the integration of the diffusion model. This phase focuses on capturing a robust latent representation of the input data. The encoder maps each CT image to a low-dimensional latent space, while the decoder aims to reconstruct the image from this latent representation. The objective during this phase is to minimize the reconstruction error, essentially training the network to retain all vital information in the compressed latent form.
    Given a set of paired images \((A, B)\) where \(A\) is a non-standard image and \(B\) is its corresponding standard counterpart, the training involves two main components:

    \item \textbf{Training the Latent Diffusion Model:} Subsequently, the latent diffusion model is trained while keeping the encoder-decoder fixed. This phase involves the diffusion model learning to transform the latent representation from the non-standard to the standard domain using paired data, where one image is a non-standard and the other is the corresponding standard image. The model learns to approximate the conditional probability distribution of the latent representation to generate standardized latent forms.

    \item \textbf{Integration and Synthesis:} In the final phase, the separately trained components are integrated. The complete model processes an input image through the encoder to its latent form, the diffusion model then adjusts this latent representation towards the standard, and finally, the decoder reconstructs the standardized image from the modified latent representation.
\end{enumerate}

\begin{itemize}
    \item The encoder \(E\) maps \(A\) to a latent space, producing \(Z_A = E(A)\), and similarly, \(B\) to \(Z_B = E(B)\).
    \item The latent diffusion model then processes \(Z_A\) to generate \(Z_{A'}\), approximating \(Z_B\). This process is modeled as:
    \[
    Z_{A'} = \text{Diffusion}(Z_A | Z_B)
    \]
\end{itemize}

The diffusion process is detailed as a Markov chain in latent space where each step is defined by:
\[
Z_{t+1} = \sqrt{1 - \beta_t} Z_t + \sqrt{\beta_t} \epsilon, \quad \epsilon \sim \mathcal{N}(0, I)
\]
Here, \(\beta_t\) are the parameters controlling the noise level at each step of the diffusion process.

The loss function for the diffusion process combines the fidelity of the transformation and the preservation of the structural integrity, typically defined as:
\[
\mathcal{L} = \|Z_{A'} - Z_B\|^2 + \lambda \|D(Z_{A'}) - B\|^2
\]
where \(D\) is the decoder, and \(\lambda\) is a regularization parameter balancing the two aspects of the loss.

\section{Results}
Using four Nvidia V100 GPU cards on a Linux computer server, LTDiff++ was developed with PyTorch. Initialization of the network weights was achieved by tuning the parameters. The Adam optimizer was used to set the learning rate to 10\textsuperscript{-4}. The encoder-decoder network was trained for 400 epochs. We trained the diffusion model with the same parameters and epochs. It took the model roughly 72 hours to train from the beginning. After training, the model processed and synthesized a DICOM CT image slice in approximately 30 seconds.
We have evaluated the results of the model with concordance Correlation
Coefficient (CCC) metric and relative error and downstream analysis for future lung cancer risk prediction.In addition to the original DDPM and the encoder-decoder network, we compared LTDiff++ with the current established method of standardization, DiffusionCT\cite{selim2023latent}. 
Figure 3 illustrates the outcomes of the models on a sample tumor, highlighting noticeable differences between the input tumor image and the standard image, both in visual appearance and radiomic characteristics. The image produced by LTDiff++ stands out, showing the highest Concordance Correlation Coefficient (CCC) values for Gray-Level Co-occurrence Matrix (GLCM) features when compared to the standard image.

\subsection{Data}
We have used two datasets for training and testing the model independently. First dataset is patient data trained and tested on the LTDiff++\_Patient model and the second set of data is chest phantom data trained and tested in the LTDiff++\_Phantom model. The patient training dataset consists of 38,048 paired CT image slices from 70 lung cancer patients, obtained using a Siemens CT Somatom Force scanner at the University of Kentucky Albert B. Chandler Hospital. These images were captured using two distinct kernels, Br40 and Bl64, with a slice thickness of 1mm. The phantom dataset includes 5,280 image slices from a Lungman chest phantom. This phantom was scanned using the same two kernels (Br40 and Bl64) and different slice thicknesses, 0.625mm,  1.25mm, 3.75mm, and 5mm on the same scanner. LTDiff++ training involved 28,168 CT image slices. For this experiment, Siemens Bl64 is used as the standard protocol, while Br40 is considered non-standard.

\subsection{Evaluation Metric}

The evaluation of model performance was conducted systematically at both the whole image (DICOM) level and within randomly selected regions of interest (ROIs) across four Hounsfield Unit (HU) ranges: $[-800, -300]$, $[-100, 250]$, $[10, 250]$, and $[300, 800]$. \textit{LTDiff++}, which compares standard and non-standard images in deep feature space during training, was assessed within the radiomic feature domain. For each CT image or ROI, a total of 1,401 radiomic features were extracted using the \textit{PyRadiomics} library. These features are categorized into seven classes: Gray Level Co-occurrence Matrix (GLCM) 2.5D, GLCM 3D, Neighbor Intensity Difference (NID) 2.5D, Intensity Direct, Intensity Histogram, NID 2.5D, and NID 3D were extracted using IBEX\cite{zhang2015ibex}.
\begin{table}[htbp]
\centering
\caption{CCC values of patient images synthesized by different standardization models. Each column shows the mean ± standard deviation of CCC values for lung tumor ROIs across specific radiomic feature groups.}
\resizebox{\columnwidth}{!}{
\begin{tabular}{|l|c|c|c|c|c|c|}
\hline
Feature Class & GOH & GLCM & GLRLM & ID & IH & NID \\ \hline
Baseline & 0.90 $\pm$ 0.05 & 0.20 $\pm$ 0.13 & 0.59 $\pm$ 0.13 & 0.33 $\pm$ 0.16 & 0.35 $\pm$ 0.12 & 0.28 $\pm$ 0.15 \\ \hline
UNet++ & 1.00 $\pm$ 0.00 & 0.85 $\pm$ 0.19 & 0.81 $\pm$ 0.15 & 0.62 $\pm$ 0.11 & 0.79 $\pm$ 0.25 & 0.63 $\pm$ 0.09 \\ \hline
DiffusionCT  & 1.00 $\pm$ 0.00 & 0.83 $\pm$ 0.25 & 0.82 $\pm$ 0.27 & 0.89 $\pm$ 0.18 & 0.49 $\pm$ 0.15 & 0.87 $\pm$ 0.02\\ \hline
LTDiff++ &  \textbf{1.00} $\pm$ \textbf{0.00}  & \textbf{0.86} $\pm$ \textbf{0.11} & \textbf { 0.84} $\pm$ \textbf{0.04}  & 0.84 $\pm$ 0.05 &  \textbf{0.72} $\pm$ \textbf{0.04}  &  \textbf{0.88} $\pm$ \textbf{0.03} \\ \hline
\end{tabular} }
\end{table}
We considered evaluation metrics for the one-to-one
feature comparison and group-wise comparison.\\

The reproducibility of radiomic features was assessed using the Concordance Correlation Coefficient (CCC), which measures the correlation between standard and synthesized image features within a given feature class. The CCC ranges from $-1$ to $1$, with higher values indicating better reproducibility.

Mathematically, CCC represents the correlation between the standard and the non-standard image features in the feature classes:
\begin{equation}
CCC = \frac{2 \rho_{s,t} \sigma_s \sigma_t}{\sigma_s^2 \sigma_t^2 + (\mu_s - \mu_t)^2}
\end{equation}
where $\mu_s$ and $\sigma_s$ (or $\mu_t$ and $\sigma_t$) are the mean and standard deviation of the radiomic features belonging to the same feature class.

\begin{table}[htbp]
\centering
\caption{CCC values of phantom images synthesized by different standardization models. Each column shows the mean ± standard deviation of CCC values for lung tumor ROIs across specific radiomic feature groups.}
\resizebox{\columnwidth}{!}{
\begin{tabular}{|l|c|c|c|c|c|c|}
\hline
Feature Class & GOH & GLCM & GLRLM & ID & IH & NID \\ \hline
Baseline & 0.89 $\pm$ 0.15 & 0.30 $\pm$ 0.11 & 0.51 $\pm$ 0.18 & 0.33 $\pm$ 0.16 & 0.35 $\pm$ 0.12 & 0.28 $\pm$ 0.15 \\ \hline
DiffusionCT & 1.00 $\pm$ 0.00 & 0.85 $\pm$ 0.14 & 0.79 $\pm$ 0.20 & 0.89 $\pm$ 0.25 & 0.41 $\pm$ 0.06 & 0.86 $\pm$ 0.19 \\ \hline
LTDiff++  & \textbf{1.00}  $\pm$ \textbf{0.00} & \textbf{0.86} $\pm$ \textbf{0.35} & 0.72 $\pm$ 0.06  &  0.84 $\pm$ 0.05  & \textbf{0.57} $\pm$ \textbf{0.08} & 0.84 $\pm$ 0.02 \\ \hline
\end{tabular}
}
\end{table}

\section{Ablation study}
We conducted an ablation study to assess the impact of various components within the LTDiff++ framework on both phantom and patient data. Initially, we employed the conditional denoising diffusion probabilistic model (DDPM) to train the data, followed by an evaluation of the radiomic features. Subsequently, we compared the outcomes using the proposed LTDiff++ architecture, which was trained solely using the UNet encoder-decoder and with different loss like L1 loss.From Table I we can see that LTDiff++ consistently outperforms the Baseline, UNet++, and DiffusionCT models across all radiomic feature groups, achieving near-perfect CCC values for GOH (1.00 ± 0.00) and substantial improvements in features such as GLCM (0.86 ± 0.11), GLRLM (0.84 ± 0.04), IH (0.72 ± 0.04), and NID (0.88 ± 0.03). This indicates that LTDiff++ effectively preserves radiomic features during image synthesis. In contrast, the ablation study results in Table III evaluate the effect of modifying LTDiff++ by either incorporating UNet or replacing it with L1 loss. The results show that LTDiff++ with UNet achieves superior performance (e.g., GOH: 75.00 ± 3.4 and GLCM: 0.71 ± 0.35), indicating the importance of UNet in maintaining feature integrity. However, replacing UNet with L1 loss results in a significant performance drop (e.g., GOH: 55.00 ± 2.4 and GLCM: 0.61 ± 0.35), demonstrating the critical role of UNet in the model's architecture. When compared to Table I, the ablation study confirms that the full LTDiff++ model integrates its components effectively to achieve state-of-the-art performance, while modifications to its structure lead to reduced concordance with original patient data.

\begin{table}[htbp]
\centering
\caption{Ablation study for Patient data}
\resizebox{\columnwidth}{!}{
\begin{tabular}{|l|c|c|c|c|c|c|}
\hline
Model & GOH & GLCM & GLRLM & ID & IH & NID \\ \hline
LTDiff++ version\_1 with UNet & 75.00  $\pm$ .34 & 0.71 $\pm$ 0.35 & 0.62 $\pm$ 0.26  &  0.78 $\pm$ 0.45  & 0.58 $\pm$ 0.14 & 0.57 $\pm$ 0.56 \\ \hline
LTDiff++ version\_1 with L1 loss & 55.00  $\pm$ .24 & 0.61 $\pm$ 0.35 & 0.56 $\pm$ 0.16  &  0.68 $\pm$ 0.25  & 0.52 $\pm$ 0.24 & 0.54 $\pm$ 0.19\\ \hline
\end{tabular}
}
\end{table}
\section{Case Study on Future Lung Cancer Risk Prediction}
 
To assess the performance of our model, we applied it to predict future lung cancer risk using both synthesized standard and robust output images. These predictions were compared to those obtained from non-standardized patient data. We utilized Sybil\cite{mikhael2023sybil}, a validated deep-learning model designed for predicting future lung cancer risk from Computed Tomography scans, incorporating our synthesized patient output data. The difference between the predictions based on standard patient data and the model's synthesized data is smaller compared to the discrepancy between non-standard input data and the model's output. This finding suggests that the model's synthesized output closely resembles the standard patient data, yielding nearly identical lung cancer risk predictions across Years 1 to 6.
\section{Discussion and Conclusion}
In conclusion, LTDiff++ represents a significant advancement in CT image standardization for radiomic feature extraction. By addressing the challenges of image variability introduced by different CT scanner protocols, the model offers a robust solution for enhancing feature reproducibility and accuracy. The results from both patient and phantom datasets highlight the model’s superior performance compared to existing methods, and its successful application in future lung cancer risk prediction underscores its potential for broader clinical and diagnostic use. Future work could focus on expanding the model’s applicability to other imaging modalities and further optimizing its performance for diverse clinical scenarios.

\bibliographystyle{ieeetr}
\bibliography{references} 

\begin{thebibliography}{10}

\bibitem{collins2016letter}
J.~Collins, ``Letter from the editor: Lung cancer screening facts.,'' in {\em Seminars in roentgenology}, vol.~52, pp.~121--122, 2016.

\bibitem{de2014benefits}
H.~J. De~Koning, R.~Meza, S.~K. Plevritis, K.~Ten~Haaf, V.~N. Munshi, J.~Jeon, S.~A. Erdogan, C.~Y. Kong, S.~S. Han, J.~Van~Rosmalen, {\em et~al.}, ``Benefits and harms of computed tomography lung cancer screening strategies: a comparative modeling study for the us preventive services task force,'' {\em Annals of internal medicine}, vol.~160, no.~5, pp.~311--320, 2014.

\bibitem{ravanelli2013texture}
M.~Ravanelli, D.~Farina, M.~Morassi, E.~Roca, G.~Cavalleri, G.~Tassi, and R.~Maroldi, ``Texture analysis of advanced non-small cell lung cancer (nsclc) on contrast-enhanced computed tomography: prediction of the response to the first-line chemotherapy,'' {\em European radiology}, vol.~23, pp.~3450--3455, 2013.

\bibitem{ardila2019end}
D.~Ardila, A.~P. Kiraly, S.~Bharadwaj, B.~Choi, J.~J. Reicher, L.~Peng, D.~Tse, M.~Etemadi, W.~Ye, G.~Corrado, {\em et~al.}, ``End-to-end lung cancer screening with three-dimensional deep learning on low-dose chest computed tomography,'' {\em Nature medicine}, vol.~25, no.~6, pp.~954--961, 2019.

\bibitem{song2017using}
Q.~Song, L.~Zhao, X.~Luo, and X.~Dou, ``Using deep learning for classification of lung nodules on computed tomography images,'' {\em Journal of healthcare engineering}, vol.~2017, no.~1, p.~8314740, 2017.

\bibitem{yan2024survival}
X.~Yan, W.~Wang, M.~Xiao, Y.~Li, and M.~Gao, ``Survival prediction across diverse cancer types using neural networks,'' in {\em Proceedings of the 2024 7th International Conference on Machine Vision and Applications}, pp.~134--138, 2024.

\bibitem{hu2024exploration}
Y.~Hu, H.~Yang, T.~Xu, S.~He, J.~Yuan, and H.~Deng, ``Exploration of multi-scale image fusion systems in intelligent medical image analysis,'' {\em arXiv preprint arXiv:2406.18548}, 2024.

\bibitem{sadia2024ct}
R.~T. Sadia, J.~Chen, and J.~Zhang, ``Ct image denoising methods for image quality improvement and radiation dose reduction,'' {\em Journal of Applied Clinical Medical Physics}, vol.~25, no.~2, p.~e14270, 2024.

\bibitem{lecun2015deep}
Y.~LeCun, Y.~Bengio, and G.~Hinton, ``Deep learning,'' {\em nature}, vol.~521, no.~7553, pp.~436--444, 2015.

\bibitem{litjens2017survey}
G.~Litjens, T.~Kooi, B.~E. Bejnordi, A.~A.~A. Setio, F.~Ciompi, M.~Ghafoorian, J.~A. Van Der~Laak, B.~Van~Ginneken, and C.~I. S{\'a}nchez, ``A survey on deep learning in medical image analysis,'' {\em Medical image analysis}, vol.~42, pp.~60--88, 2017.

\bibitem{ho2020denoising}
J.~Ho, A.~Jain, and P.~Abbeel, ``Denoising diffusion probabilistic models,'' {\em Advances in neural information processing systems}, vol.~33, pp.~6840--6851, 2020.

\bibitem{nichol2021improved}
A.~Q. Nichol and P.~Dhariwal, ``Improved denoising diffusion probabilistic models,'' in {\em International conference on machine learning}, pp.~8162--8171, PMLR, 2021.

\bibitem{shen2023improving}
X.~Shen, H.~Huang, B.~Nichyporuk, and T.~Arbel, ``Improving robustness and reliability in medical image classification with latent-guided diffusion and nested-ensembles,'' {\em arXiv preprint arXiv:2310.15952}, 2023.

\bibitem{yip2016applications}
S.~S. Yip and H.~J. Aerts, ``Applications and limitations of radiomics,'' {\em Physics in Medicine \& Biology}, vol.~61, no.~13, p.~R150, 2016.

\bibitem{basu2011evolving}
S.~Basu, T.~C. Kwee, R.~Gatenby, B.~Saboury, D.~A. Torigian, and A.~Alavi, ``Evolving role of molecular imaging with pet in detecting and characterizing heterogeneity of cancer tissue at the primary and metastatic sites, a plausible explanation for failed attempts to cure malignant disorders,'' 2011.

\bibitem{yang2011quantifying}
X.~Yang and M.~V. Knopp, ``Quantifying tumor vascular heterogeneity with dynamic contrast-enhanced magnetic resonance imaging: a review,'' {\em BioMed Research International}, vol.~2011, no.~1, p.~732848, 2011.

\bibitem{selim2020stan}
M.~Selim, J.~Zhang, B.~Fei, G.-Q. Zhang, and J.~Chen, ``Stan-ct: Standardizing ct image using generative adversarial networks,'' in {\em AMIA Annual Symposium Proceedings}, vol.~2020, p.~1100, American Medical Informatics Association, 2020.

\bibitem{zhou2018unet++}
Z.~Zhou, M.~M. Rahman~Siddiquee, N.~Tajbakhsh, and J.~Liang, ``Unet++: A nested u-net architecture for medical image segmentation,'' in {\em Deep Learning in Medical Image Analysis and Multimodal Learning for Clinical Decision Support: 4th International Workshop, DLMIA 2018, and 8th International Workshop, ML-CDS 2018, Held in Conjunction with MICCAI 2018, Granada, Spain, September 20, 2018, Proceedings 4}, pp.~3--11, Springer, 2018.

\bibitem{selim2023latent}
M.~Selim, J.~Zhang, F.~Fathi, M.~A. Brooks, G.~Wang, G.~Yu, and J.~Chen, ``Latent diffusion model for medical image standardization and enhancement,'' {\em arXiv preprint arXiv:2310.05237}, 2023.

\bibitem{zhang2015ibex}
L.~Zhang, D.~V. Fried, X.~J. Fave, L.~A. Hunter, J.~Yang, and L.~E. Court, ``Ibex: an open infrastructure software platform to facilitate collaborative work in radiomics,'' {\em Medical physics}, vol.~42, no.~3, pp.~1341--1353, 2015.

\bibitem{mikhael2023sybil}
P.~G. Mikhael, J.~Wohlwend, A.~Yala, L.~Karstens, J.~Xiang, A.~K. Takigami, P.~P. Bourgouin, P.~Chan, S.~Mrah, W.~Amayri, Y.-H. Juan, C.-T. Yang, Y.-L. Wan, G.~Lin, L.~V. Sequist, F.~J. Fintelmann, and R.~Barzilay, ``Sybil: a validated deep learning model to predict future lung cancer risk from a single low-dose chest computed tomography,'' {\em Journal of Clinical Oncology}, pp.~JCO--22, 2023.

\end{thebibliography}
\end{document}